\documentstyle[12pt,epsf]{elsart}

\begin{document}
\begin{frontmatter}
\title{J/$\psi$ Suppression in Heavy Ion Collisions\\ at the CERN SPS}  
\author[DEK]{D.~E.~Kahana} \and \author[SHK]{S.~H.~Kahana}
\address[SHK]{Physics Department, Brookhaven National Laboratory,
Upton, NY 11973}
\address[DEK]{31 Pembrook Drive, Stony Brook, NY 11790}

\begin{abstract}
We reexamine the production of J$/\psi$ and other charmonium states for a
variety of target-projectile choices at the SPS, in particular for the
interesting comparison between S+U at $200$ GeV/c and Pb+Pb at $158$ GeV/c as
observed in the experiments NA38 and NA50 respectively. The result is a
description of the NA38 and NA50 data in terms of a conventional, 
quasi-hadronic picture.  The apparently anomalous suppression found in the 
most massive Pb+Pb system arises in the present simulation from three
sources: destruction in the initial nucleon-nucleon cascade phase, use of
coupled channels to exploit the larger breakup in the less bound $\chi^{i}$
and $\psi'$ states, and comover interaction in the final low energy phase.
\end{abstract}
\end{frontmatter}

\bigskip

\section{Introduction}  
The possible use of J$/\psi$ suppression as a signal of unusual behaviour in
relativistic ion collisions, first suggested by Matsui and Satz
\cite{Matsui}, has attracted considerable experimental and theoretical study.
Great interest has attached to the results obtained by the NA50 collaboration
for charmonium production in Pb+Pb collisions at $158$ GeV/c: to the early
findings presented at the Quark Matter 1996 meeting \cite{NA50a} as well as
to the startling data later released at RHIC'97 \cite{NA50b}.  The success of
Glauber-like calculations of J/$\psi$ production and breakup in the p+A and
S+U \cite{Huefner,Kharzeev,Gavin} systems, coupled with a failure of Glauber to
provide an equally good description of the apparently accelerated absorption
in Pb+Pb has been widely interpreted \cite{NA50a,NA50b,Kharzeev} as a signal
of QCD plasma creation in these collisions. The very sharp behaviour of the
J/$\psi$ yield as a function of transverse energy $E_t$ seen in the later
experiment \cite{NA50b} has especially attracted attention. 

We attempt to retrace this ground theoretically \cite{LANLBNL}, employing a
new, two phase cascade approach, described in detail elsewhere
\cite{LUCIFERI,LUCIFERII}, combined with a variation of the Satz-Kharzeev
model for production and annihilation of charmonium in the initial baryonic
collisions. This modeling described below, allows the coupled-channel aspect
of the hidden charm spectroscopy, $\left\lbrace \psi, \chi^{i},
\psi'\right\rbrace$ to play a more central role.  In this first application,
we include partons in a minimal fashion, to describe for example Drell-Yan
production. Hence we are testing a `purely' hadronic description of the
anomalous Pb+Pb measurements.

It has been pointed out that a hadronic picture might succeed \cite{Gavin}
without invoking quark-gluon plasma (QGP) creation, if at least part of the
seemingly anomalous suppression in Pb+Pb could be produced by comover
annihilation, {\it i.e.\ }by interactions of the J/$\psi$ with secondary
mesons generated in the ion-ion collision. The second phase in LUCIFER II,
which is a low energy cascade, perforce includes the effect of J/$\psi$
destruction through such comover rescattering.

\begin{figure}[htb]
\begin{minipage}[t]{65mm}
	   \epsfysize = 65 mm
	   \centerline{ \epsfbox{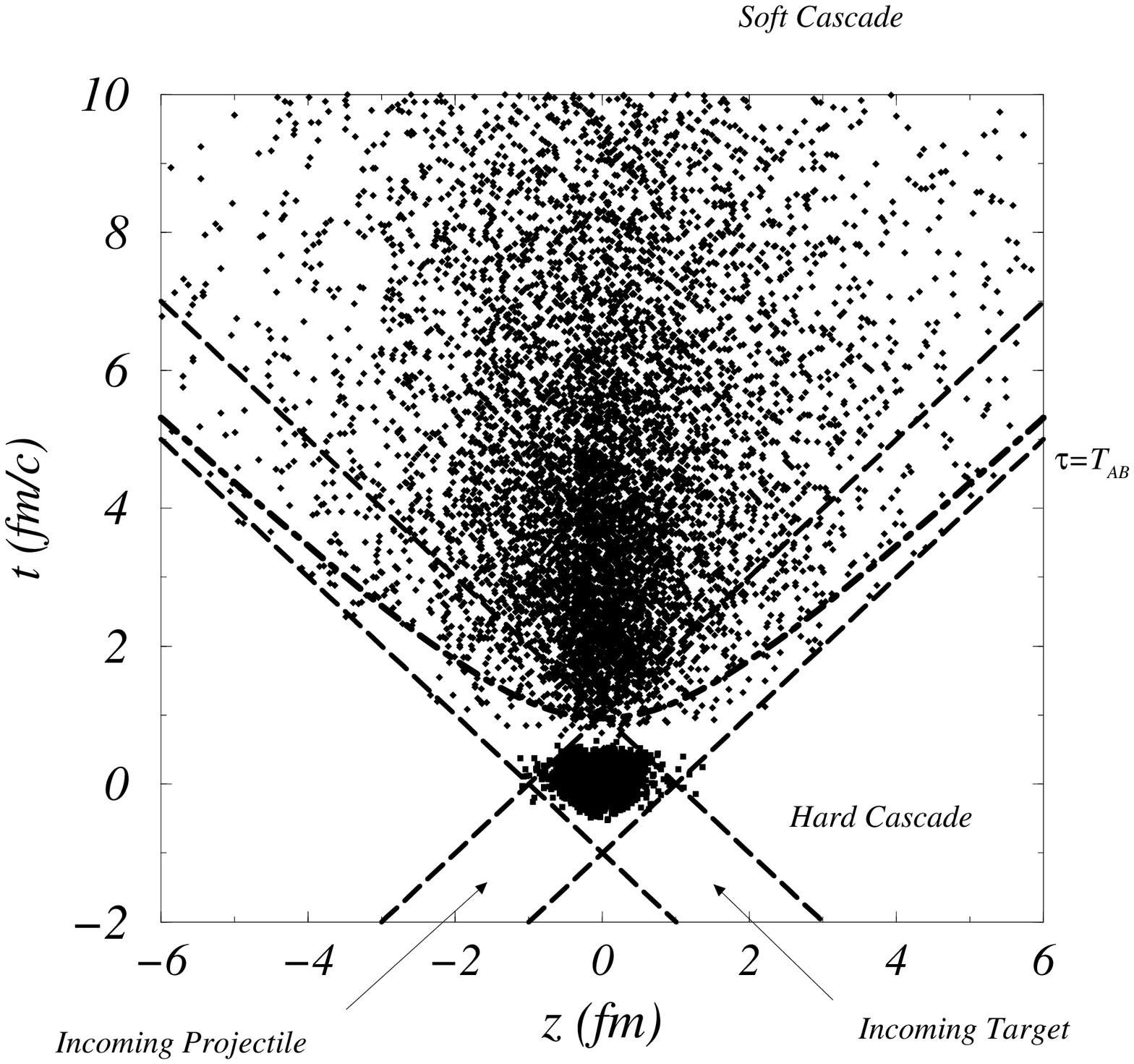} }
\caption[]{Time evolution of the ion-ion collision. The space-time distribution 
of collisions and decays in hard and soft cascades is shown for the minimum
bias Pb+Pb system.}
\label{fig:collisions}
\end{minipage}
\hskip 0.25truein
\begin{minipage}[t]{65mm}
	   \epsfysize = 65 mm
	   \centerline{ \epsfbox{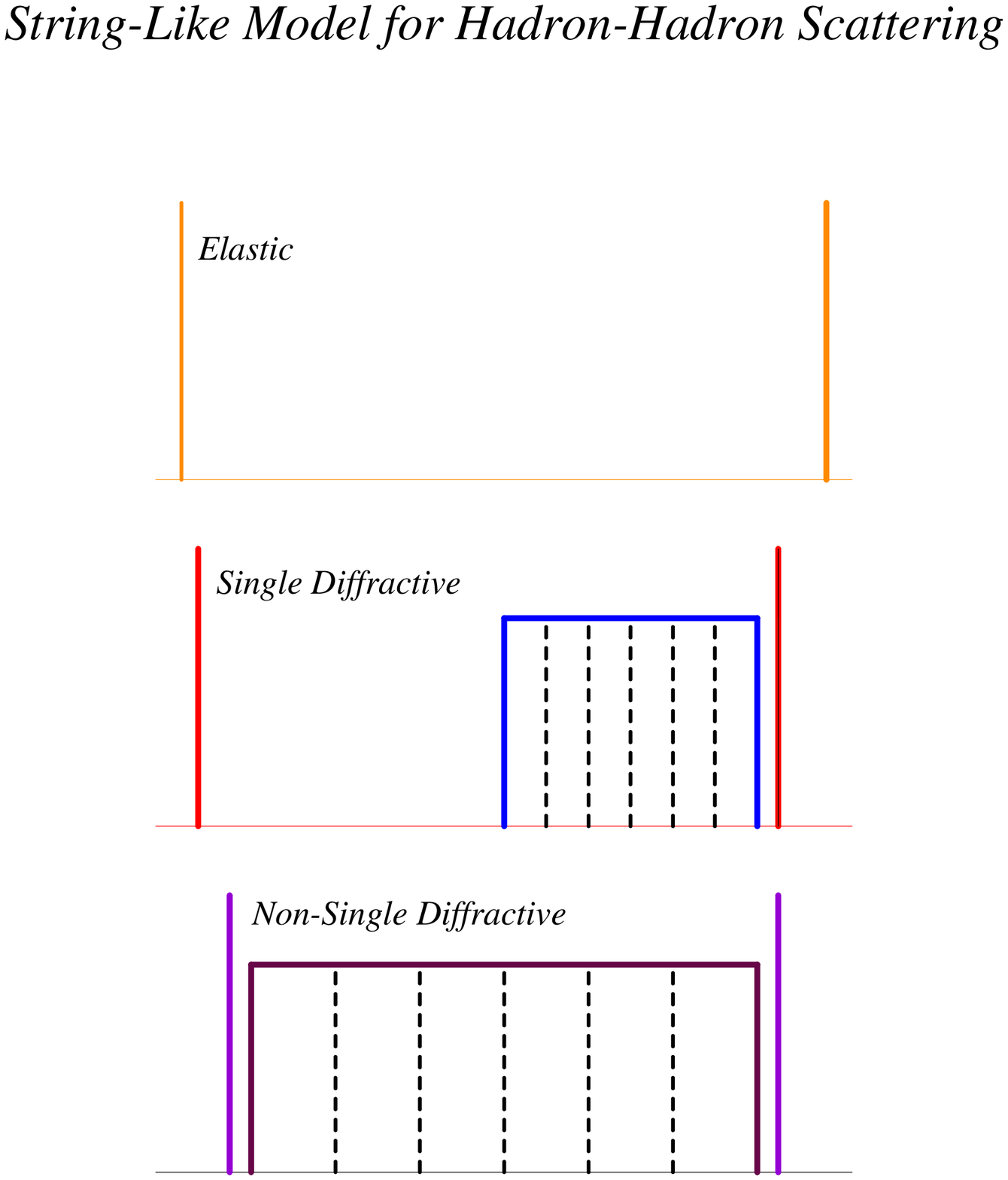} }
\caption[]{Shown are graphic representations of the elements of the model for
the elementary hadron-hadron collision: elastic, single diffractive (SD) and
non-single diffractive (NSD) scattering.}
\label{fig:basic_pp}
\end{minipage}
\end{figure}

In LUCIFER II we have attempted to separate hard and soft processes by time
scale, see Fig(\ref{fig:collisions}), so as to permit partonic and hadronic
cascading to be joined naturally, in a modular fashion. The separation is
effected through the use of a short time scale, automatically present at high
energies: the time $T_{AB}$ taken for the two interacting nuclei A and B to
traverse each other in the global collision frame. The uncertainty principle
allows hard interactions involving sufficiently high energy-momentum
transfer, {\it i.e.\ } for $Q^{-1} \le T_{AB}$, to take place in the first
and very rapid cascade. Soft processes involving low tranverse momentum are
not completed until later. Thus in the initial fast cascading the nucleons
{\it lose no energy} but are still aware of the number and nature of the
two-particle collisions they have undergone.

Specifically, the method \cite{LUCIFERII} consists of running the cascade in
two stages. The first is a high energy fast-time mode in which collision
histories are recorded and fast processes (Drell-Yan and charmonium
production) are allowed to occur. Using the entire space-time and
energy-momentum history of this stage, a reinitialisation of the cascade is
performed using elementary hadron-hadron data as a strict guide. The final
positions and momenta of baryons in the first phase, and the number of
collisions they suffer are recorded and used to generate produced mesons
together with their initial momentum and space-time coordinates. In the
initial ion-ion collision the interacting nucleon paths are almost along
light-cones. The second cascade begins at $T_{AB}$, the
time of the last nucleon-nucleon collision, with initial conditions specified
by the reinitialisation, but no secondary interactions are allowed until a formation
time for produced mesons has passed.  The participants in the second phase
are generic mesons, thought of as of $q\bar q$-like in character with masses
centered near $M_{q\bar q}=700$ MeV and in the range $M_{q\bar q} \sim
0.3-1.1$ GeV. Generic baryons consisting of $qqq$ are also included and are
excited to rather light masses, $M_{qqq}\sim 0.94-2.0$ GeV \cite{LUCIFERII}.
All the generic hadrons decay {\it via} sequential pion emission.  Normal
stable mesons and baryons are also present, and terminate the decay chains.
See Fig(\ref{fig:basic_pp}) for a pictorial description of the model for
basic hadron-hadron scattering.

Kharzeev and Satz \cite{Kharzeev} employ a model based on hadronic Glauber
theory describing production and breakup of the J/$\psi$ in ion-ion
collisions, to demonstrate that such a picture cannot account for the degree
of suppression seen in Pb+Pb collisions at the SPS. Reasoning similarly, we
can make a close comparison of our treatment with their work. The required
initial production of a $c \bar c$ pair is handled within an effective
hadronic formulation both in our work and in that of Kharzeev, {\it et.al.\
}. There are, naturally, specific and important differences between Glauber
theory and a cascade model, and it is partly these differences which permit
the so-called anomalous suppression in Pb+Pb to be explained within a purely
hadronic framework. 

The overall degree of suppression in Pb+Pb, insofar as it differs from
earlier work \cite{Kharzeev,Gavin}, results from a combination of effects;
these are baryonic, coupled channel and comover in kind, with substantial
contributions arising from both phases of the cascade. There are potential
unknown variables: the production and dynamic time evolution of each
charmonium state, the breakup probabilities against both baryons and mesons,
the density of secondary mesons. This last is to a large extent predicted by
the cascade, which must agree with actual inclusive final state meson and
baryon distributions.  There also exist constraints on the basic charmonium
variables. The production is in principle determined in elementary
nucleon-nucleon collisions, the baryonic breakup in p+A collisions. The
$\psi \pi$ breakup cross-sections are not known directly from any
measurements. 

We leave in this conference proceeding report the details of the cascade
architecture to the perusal of a previous publication \cite{LUCIFERII}.

\section{Coupled Channel Model for Charmonium}

\begin{figure}[htb]
\begin{minipage}[t]{65mm}
	   \epsfysize = 65 mm
	   \centerline{ \epsfbox{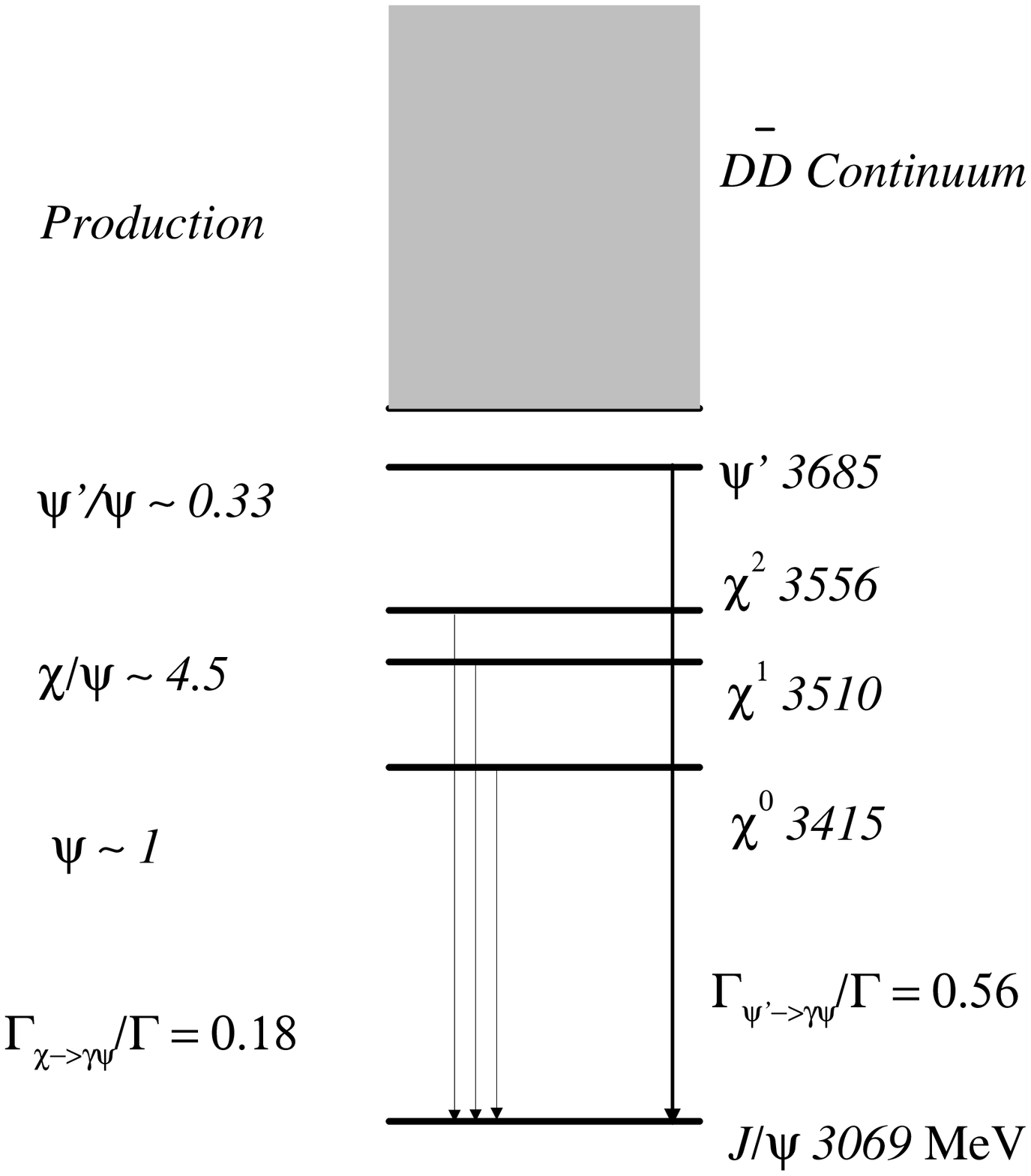} }
\caption[]{Charmonium spectroscopy including higher mass states which are
significantly produced in $pp$ and which feed strongly to the J$\psi$.
Electromagnetic and hadronic decays of $\chi^i$ (a weighted average) and
$\psi'$ are both included in the indicated branching ratios. The production
ratios are suggested by direct measurement.} 
\label{fig:levels}
\end{minipage}
\hskip 0.25truein
\begin{minipage}[t]{65mm}
	   \epsfysize = 65 mm
	   \centerline{ \epsfbox{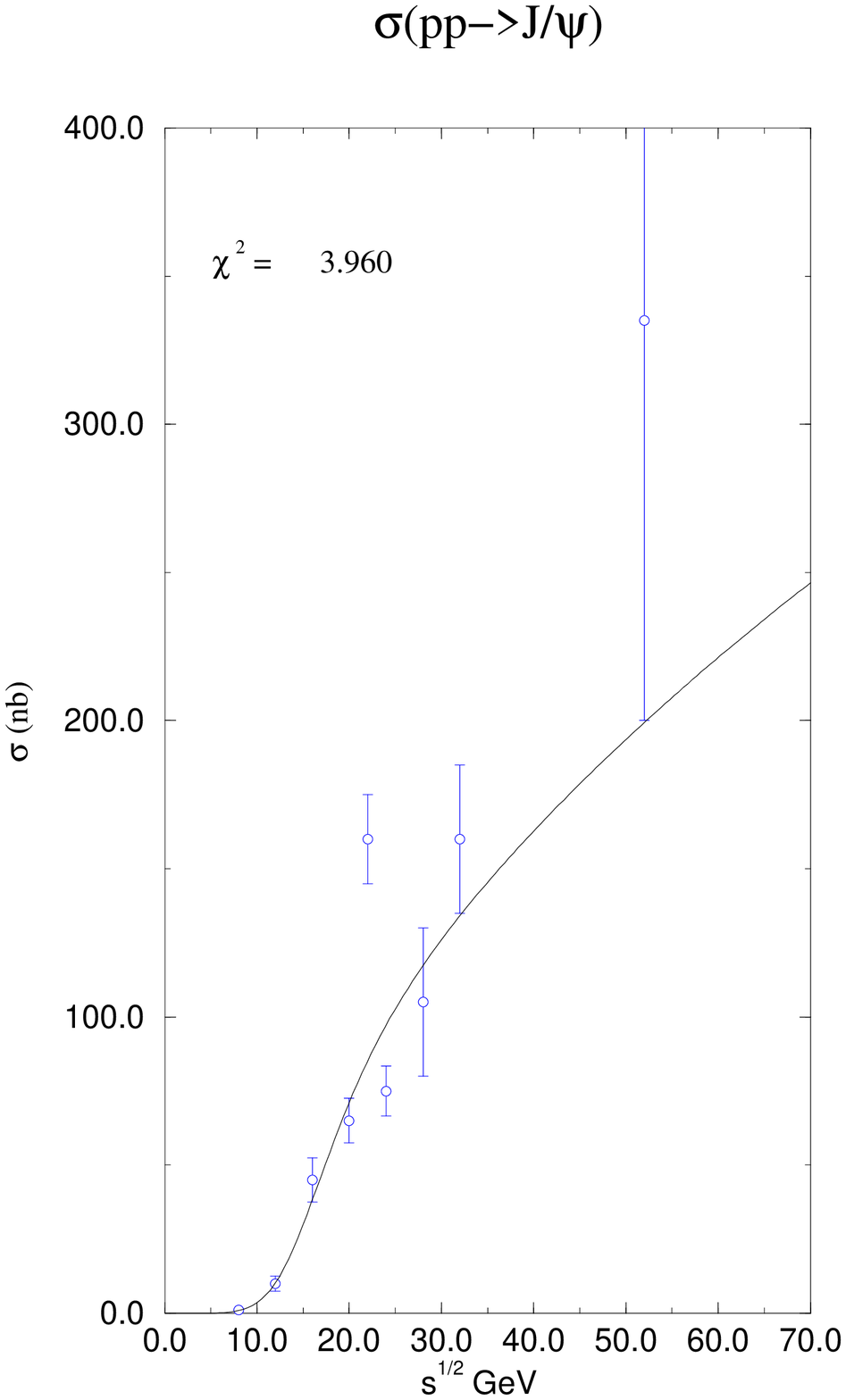} }
\caption[]{Production of J/$\psi$ from pp as a function of energy. The data
appear as points and the fit as a line. The $\pi p$ cross-section is also
known, and in fact is very similar to that for $pp$, but rarely plays a role
with production generally significant only at the highest energies.}
\label{fig:jpsi-xs}
\end{minipage}
\end{figure}

The treatment of the hidden charm {$c\bar c$} mesons within a `purely'
hadronic code presents some problems, perhaps not fully solvable within the
effective hadronic treatment of such states. We do not deviate much in spirit
from the work of previous researchers \cite{Huefner,Kharzeev,Gavin}, but the
devil lies sufficiently in the details to produce some quantitative
effects. The production of charmonium mesons is almost completely limited to
that coming from nucleon-nucleon collisions at the highest energies, {\it
i.e.\ }in the initial high energy cascade, not by fiat but by the greatly
reduced collision energies in the second phase. Destruction of the charm
meson precursors, in contrast, can occur in the first baryonic phase and also
later in collisions with generic mesons and baryons in the second, low energy
phase, {\it i.e.\ }on comovers. It is in the destruction of the charmonium
states we differ most, ascribing a more direct role to the presence of the
higher mass $\chi$ and $\psi'$ mesons, for which in fact breakup is far
easier. We include in Fig(\ref{fig:levels}) a level diagram showing the
relevant charmonium states to make the picture as clear as possible.

In the actual calculations of the above cited references both production and
breakup are treated as instantaneous. There is no J/$\psi$ formation time in
the high energy phase or its near cousin, the Glauber or eikonal
modeling. Kharzeev {\it et al.\ }\cite{Kharzeev} in fact justify such a
choice by referring to microscopic production of charmed quark pairs, the
subsequent formation of a preresonant-resonant state from which all
charmonium mesons emanate, and the breakup in relatively hard scattering by
gluons radiating from nearby nucleons. 

We imagine that the primordial $c\bar c$ pairs are originally produced
essentially in plane wave states. Clearly, both singlet and octet color
states are involved. This view is reasonable given the
predominance of open charm production over hidden charm production in free
space $NN$ collisions. We further suppose that in elementary $NN$ collisions
the {$c\bar c$} pair eventually coalesces, with a state dependent
probability, into a J/$\psi$, $\psi'$ or $\chi$. The time which elapses will
be determined by the size of the bound state and the probability that a
transition occurs. The probability of formation will depend critically on the
relative momentum of the coalescing pair as well as on their spatial
separation. 

Therefore, whether one sees the early evolution of the eventual charmonium as
a preresonant state or as plane waves may be immaterial. Given the small size
of the J/$\psi$ as opposed to the much larger $\psi'$ and $\chi$, the
separation of the $c$ and $\bar c$ in the plane wave picture could equally
well serve as a distinguishing feature. What can differentiate our
calculation from the earlier models is the possibility of transitions between
charmonium states dynamically in a nuclear environment. Certainly, the
$\chi^1$, $\chi^2$ and $\chi^3$ states are produced considerably more
copiously in basic $pp$ collisions, with perhaps as high a ratio as
$\chi/\psi = 4-5$ \cite{isr-chi}, and they decay appreciably into J/$\psi$,
with branching ratios in the range $\Gamma_b/\Gamma \sim 12-25$\%
\cite{bluebook}.  The $\psi'$ also feed some $57\%$ into J/$\psi$. It follows
that one cannot ignore their presence.

This point becomes even more significant when one considers what the breakup
probabilities for the higher mass charmonium states are likely to be, either
in the fast or slow cascades. These heavier objects are considerably larger
spatially and might well have total cross-sections on baryons or mesons
proportional to the square of their colour dipole radius \cite{Huefner2}.  
One of our conclusions will be that a considerable portion
of the anomalous suppression seen in Pb+Pb, even for quite large impact
parameters, is a result of breakup in the higher charmonium states and an
extinction of their free space feeding down.

What extra parameters has our model introduced relative to other treatments?
We introduce  a reaction matrix for charmonium states $R_{ij}$ which permits
transitions between the states as well as diagonal, breakup,
elements. Unitarity constrains this matrix. The diagonal elements should be
present for other practitioners also but are not, in general, since only one
preresonant charmonium state is usually considered.   The quantitative
reaction matrix $R_{ij}$ is specified in ensuing sections, but is in this
work  limited to diagonal. 

The formation time for secondary mesons, $\tau_f$, controls the density of
comovers at the onset of charmonium destruction. A reasonable choice for this
parameter is $\tau_f\sim 0.5-1$ fm/c, in fact consistent with the production
of $\pi$ mesons at the SPS \cite{LUCIFERII,NA49,NA35}.  The effective
formation time is actually somewhat longer, since it is increased by the
duration of the fast cascade, {\it i.e.\ } $\tau_{eff}=\tau_f+T_{AB}/2$.

The energy dependence of the elementary J/$\psi$ production cross-sections is
shown in Fig(\ref{fig:jpsi-xs}).  The sharp dependence of $\sigma_{J/\psi}$
on energy near the SPS values $\sqrt{s}= 17-20$ GeV implies that virtually
all production occurs in the high energy phase. 
 
\section{Drell-Yan}

The high energy phase, designed to record the initial interactions of the
nucleons in the two colliding nuclei also provides the basis for our estimate
of massive dilepton production. {\it i.e.\ }Drell-Yan, an important side of
the quandary we faced at the start. We limit ourselves to the canonical FNAL
\cite{E772} p+A measurement at 800 GeV/c, but in fact the method of
calculation guarantees agreement with the lower energy p+A and A+B collected
by NA50\cite{NA50a}.  Drell-Yan is generally considered to be calculable
perturbatively for dilepton pairs with masses in excess of $M_{\mu\mu}=4$
GeV. Production in the short time defined by such masses proceeds without
energy loss and leads to the linear $A$-dependence shown in 
Reference \cite{LUCIFERII}. 

Any cascade which does not correctly describe the $A$ dependence of
Drell-Yan is in danger of producing spurious charmonium suppression by means
of premature energy loss. This point cannot be emphasized enough.

\section{Charmonium Suppression in Nuclear Collisions}

\subsection{Minimum Bias: p+A and comparison to Glauber}

We begin with the suppression in p+A for which comovers play little role.
Even here, the first stage high energy cascade does not suffice for
an accurate description, some of the suppression on baryons occurs only in
the second stage, as slow J/$\psi$'s emerging from the interaction region are
caught by nucleons, or interact at low energy in the target.  To facilitate a
comparison with the cascade we have made our own calculations with the
Glauber formalism \cite{LANLBNL,Kharzeev}.


It is instructive to extend this comparison to A+A collisions to demonstrate
that even Glauber does not reproduce the canonical power law, implied in the
experimental descriptions \cite{NA50a,NA38,NA49}, supposedly arising from
purely baryonic breakup.  These results are displayed in
Fig(\ref{fig:Glauber}) for the J/$\psi$.  The absorption cross-section 
taken to reproduce the p+A observations at $800$ GeV/c \cite{E772},
$\sigma_{abs}\sim 7.0$ mb , is equally successful for the lower SPS energies.

A second comparison can be made using the coupled channel modeling, which we
employ in the full LUCIFER calculations below.  The relative production of
the different charmonium states is taken so as to reproduce the pp data from
the ISR \cite{isr-chi} for the $\chi$ to J/$\psi$ ratio, {\it i.e.\
}$\chi/\psi\sim 4.5$, and for appropriate $\psi'$ production
\cite{pp-psip}.  Our final results are rather insensitive to a choice in this
range, since a decrease could easily be compensated by a small transition 
matrix element between J/$\psi$ and $\chi$.
 
A lesson, key to our development, is that the coupled channel model
reproduces the Glauber result for J/$\psi$, using a smaller direct breakup
cross-section, $\sigma_{abs}(J/\psi)= 5-6.0$ mb, but including indirect
destruction via the considerably larger
$\sigma_{abs}(\chi)=3\,\sigma_{abs}(J/\psi)$ for $\chi$ and perhaps higher
for $\psi'$. The increased spatial sizes of the higher states strongly
support the use of larger absorption cross-sections. One observes that pure
Glauber theory and the first stage nucleon cascade, both produce deviations
downward from any power law, and thus gathers there is a little bit of 
`anomalous' suppression even in a bare bones, no comover, theory.

\subsection{Suppression in A+B Collisions}

To complete the picture one must allow the soft cascade to go forward
for ion collisions where the production of mesons becomes 
significant. There are two sets of data to be considered: first, minimum bias
J/$\psi$ cross-sections as a function of the product $A\times B$ of nuclear
atomic numbers, and second the ratio of J/$\psi$ yield to Drell-Yan yield as
a function of centrality, or more specifically transverse energy $E_t$.

\begin{figure}[htb]
\begin{minipage}[t]{65mm}
	   \epsfysize = 65 mm
	   \centerline{ \epsfbox{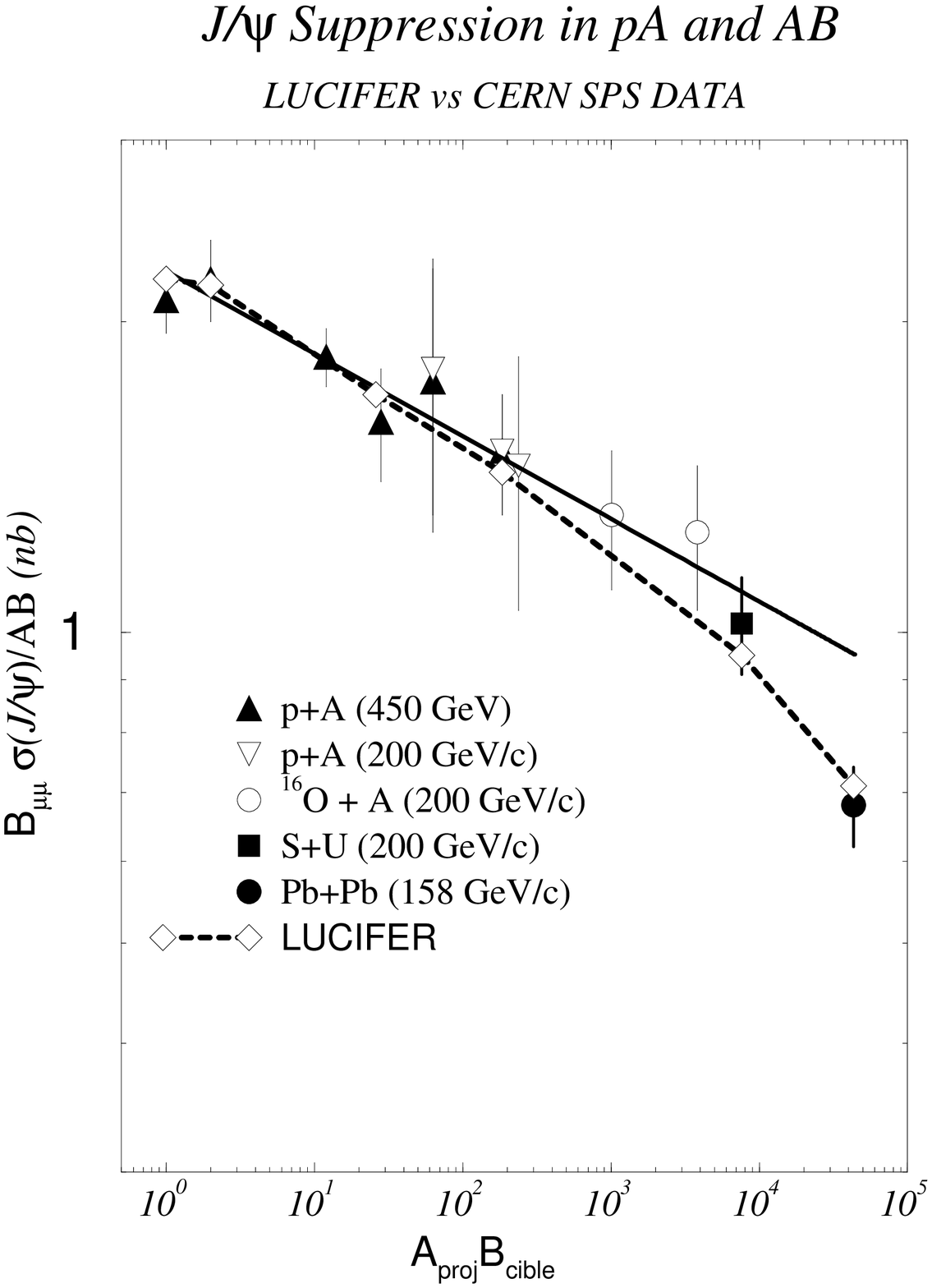} }
\caption[]{The whole range of yields for J/$\psi$ from $pp$ and p+D to 
Pb+Pb calculated in the cascade and compared to SPS measurements at various
energies. The absolute theoretical values are obtained by normalisation to 
nucleon-nucleon.}
\label{fig:jpsi-minbias}
\end{minipage}
\hskip 0.25truein
\begin{minipage}[t]{65mm}
	   \epsfysize = 65 mm
	   \centerline{ \epsfbox{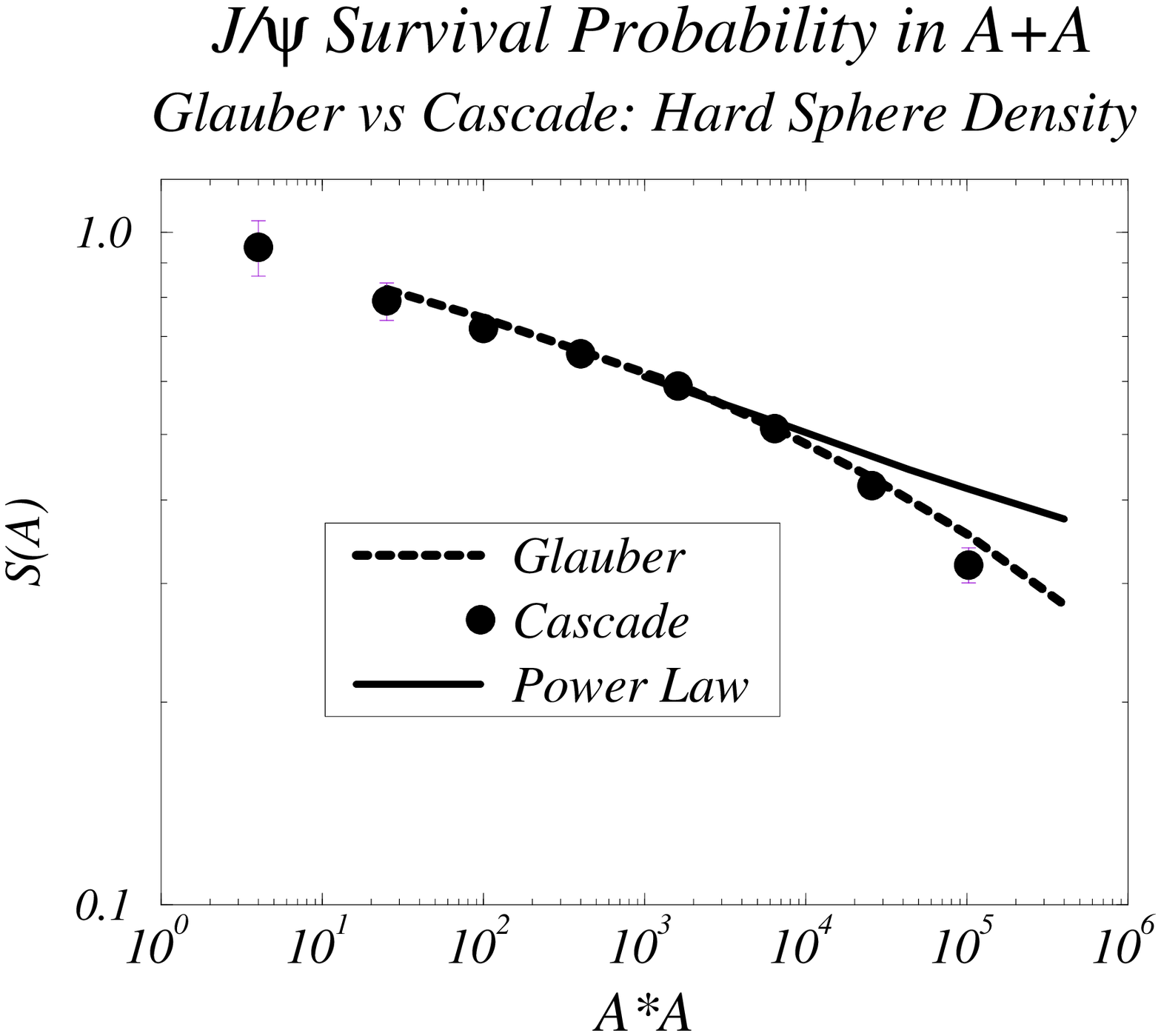} }
\caption[]{Comparison for A+A between Glauber and cascade, the latter in a
purely J/$\psi$ mode and both calculations employ $\sigma_br=7.3$ mb. The
deviation from a power law is apparent for large A$\times$A. A hard sphere form is
used for the nuclear density.} 
\label{fig:Glauber}
\end{minipage}
\end{figure}

Our results for minimum bias are displayed for the combined effect of both
cascade phases in Fig(\ref{fig:jpsi-minbias}). The anomalous suppression in
Pb+Pb is well reproduced by the totality of our two step, but otherwise
conventional, hadronic dynamics.  Part of the additional suppression in Pb+Pb
relative to S+U already arises from the high energy cascade, coming from the
increased $\chi$ and $\psi'$ breakup in the more massive nuclear
collision. But a considerable differential suppression arises from comovers,
{\it both mesonic and hadronic}, some $40\%$ of the difference between S+U
and Pb+Pb. Part of the anomaly however, is perhaps illusory in view of the
`curving down' seen for large in Fig(\ref{fig:Glauber}). 

\begin{figure}[htb]
\begin{minipage}[t]{65mm}
	   \epsfysize = 65 mm
	   \centerline{ \epsfbox{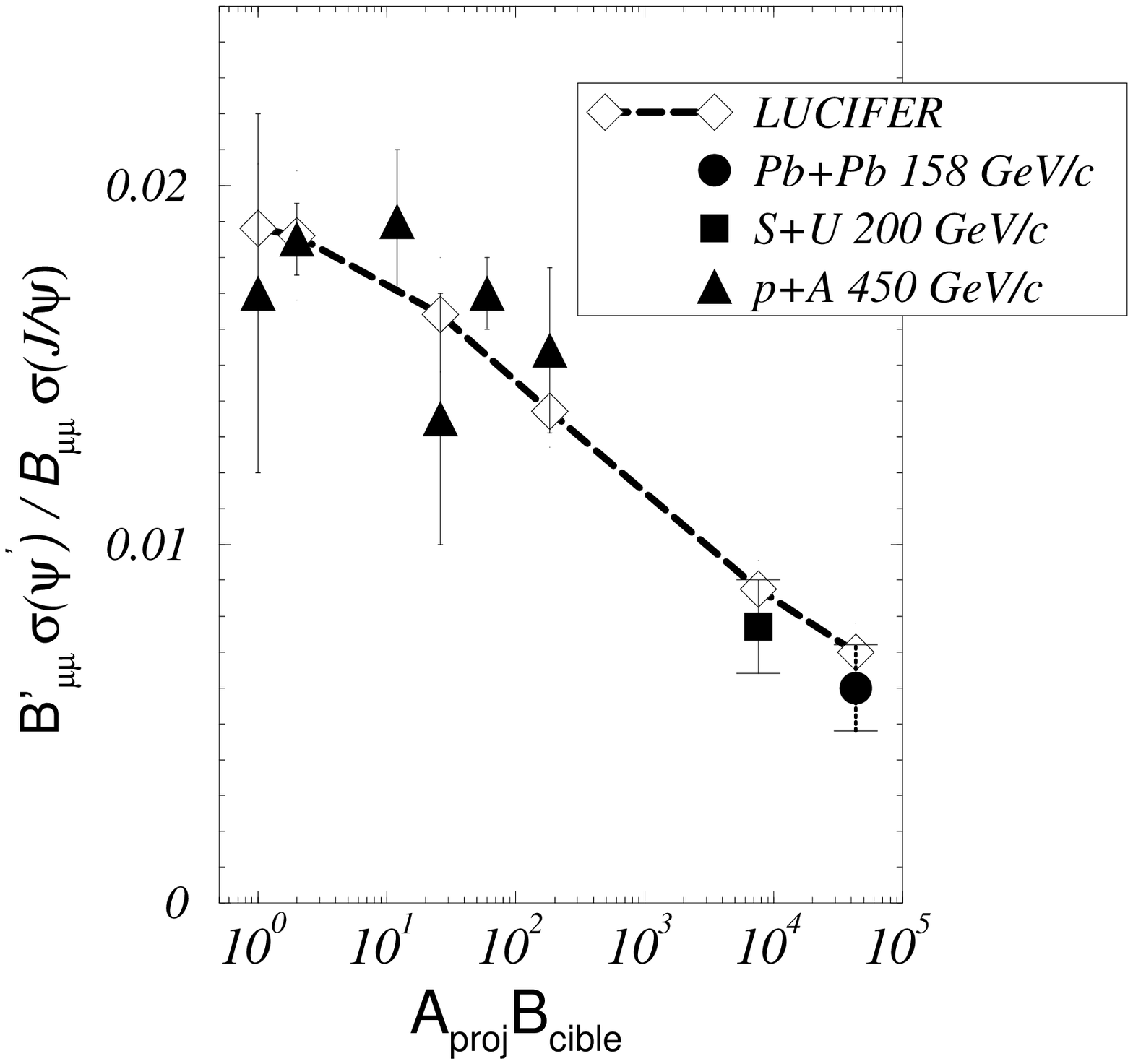} }
\caption[]{Comparison of experiment vs simulation for $\psi'$. 
The Pb+Pb data from NA50 was rescaled to 200 GeV/c by the collaboration. 
The S+U data is taken from NA38. The cascade calculations, again normalised
to nucleon-nucleon, reproduce the observed behaviour for p+A and the sharp
drop in the $\psi'$ to J/$\psi$ branching ratios for the massive nuclear collisions.}
\label{fig:psip-minbias}
\end{minipage}
\hskip 0.25truein
\begin{minipage}[t]{65mm}
	   \epsfysize = 65 mm
	   \centerline{ \epsfbox{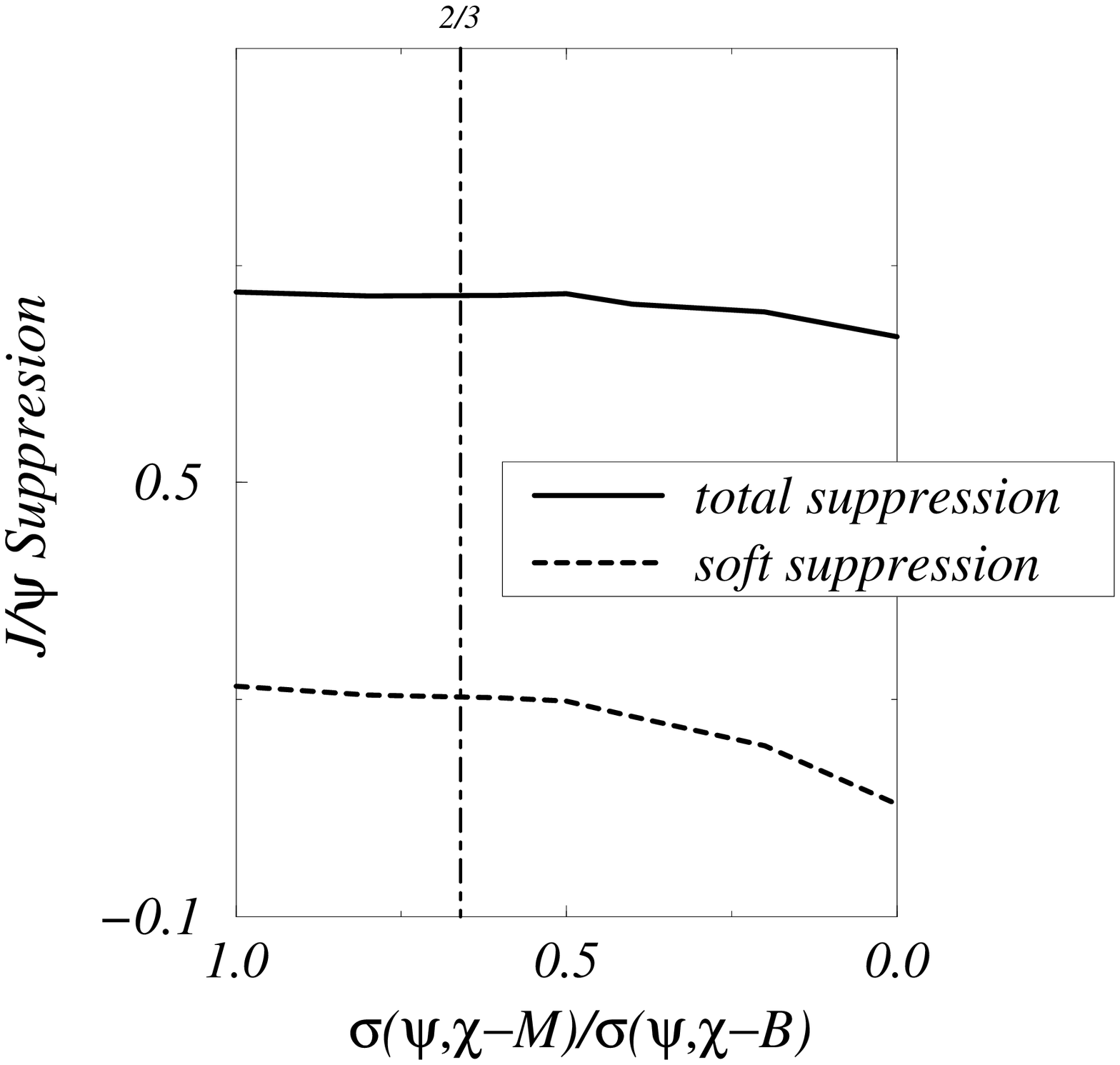} }
\caption[]{Variation of J/$\psi$ suppression with the charmonium-meson
cross-sections. We use 2/3 as the ratio to charmonium-baryon for the
calculations in the paper. The small variation with cross-section size
is surprising, and indicates a saturation is taking place in the charmonium
suppression.}
\label{fig:comovers}
\end{minipage}
\end{figure}

The calculated minimum bias $\psi'$ suppression is compared to data in
Fig(\ref{fig:psip-minbias}). The strong drop occasioned by the large increase
from p+W to S+U or Pb+Pb is clearly present in the theory. As
is evident in this figure  the $\psi'$ breakup strength inferred from p+A 
proves sufficient for both S+U and Pb+Pb.

The breakup cross-sections in these simulations are $6.6$, $20.0$ and $25.2$
mb for the $\psi$, $\chi^{i}$ and $\psi'$ respectively. These represent
absorption in charmonium-baryon collisions, and are reduced by the
constituent quark factor $2/3$ in $\psi$-meson. Variation of these
meson-meson cross-sections upwards to full equality with charmonium-baryon
leads to $\sim 0.5\%$ change in the overall J/$\psi$ suppression for
Pb+Pb (see Fig(\ref{fig:comovers})). 

This perhaps surprising non-linearity for the J/$\psi$ interaction arises
because in our model this state is intrinsically tied up with the higher
states.  Introducing off diagonal elements $R_{ij}$ would produce a family of
solutions. We have left well enough alone; the present few modeling parameters,
now mostly determined independently of S+U or Pb+Pb data, surely having
produced already an adequate description of observation, and any attempt to
force a statistically {\it accurate}  theory of the NA50 data, though 
possible, is not justified in such schematic models.

The survival probabilities for the J/$\psi$ in S+U are $0.50$ and $0.67$
in the hard and soft cascades respectively. The same figures for Pb+Pb
are $0.42$ and $0.775$.  

\subsection{Centrality: Dependence on Transverse Energy}

\begin{figure}[htb]
\begin{minipage}[t]{65mm}
	   \epsfysize = 65 mm
	   \centerline{ \epsfbox{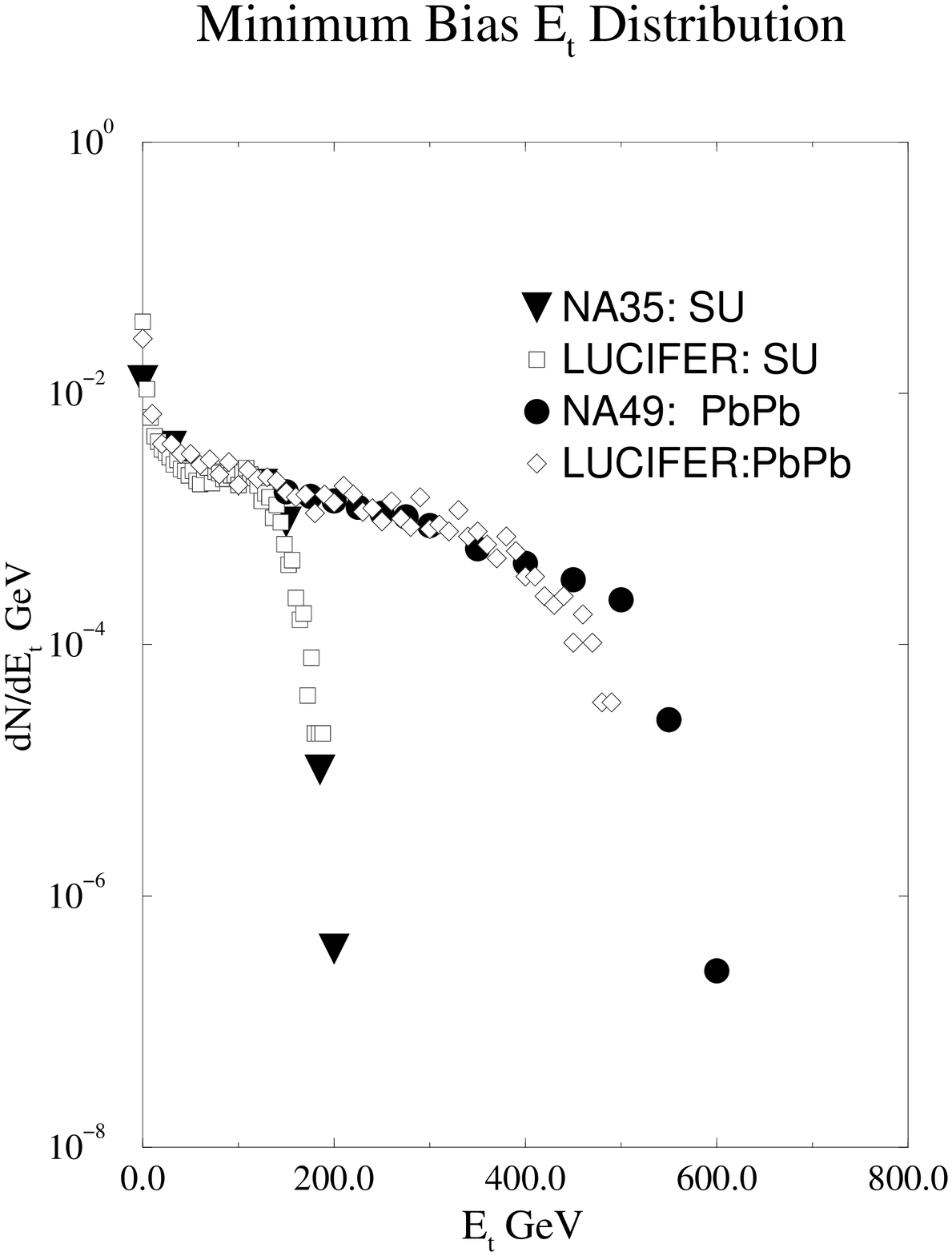} }
\caption[]{Transverse energy distributions from LUCIFER compared to 
experiment, for NA49. A broader range of pseudo-rapidity has been
used and both neutral and charged energies are present. The
calulation has underestimated the measured endpoint by some
10-15\%, which must be taken into account in considering
charm suppression.}
\label{fig:ET-NA49}
\end{minipage}
\hskip 0.25truein
\begin{minipage}[t]{65mm}
	   \epsfysize = 65 mm
	   \centerline{ \epsfbox{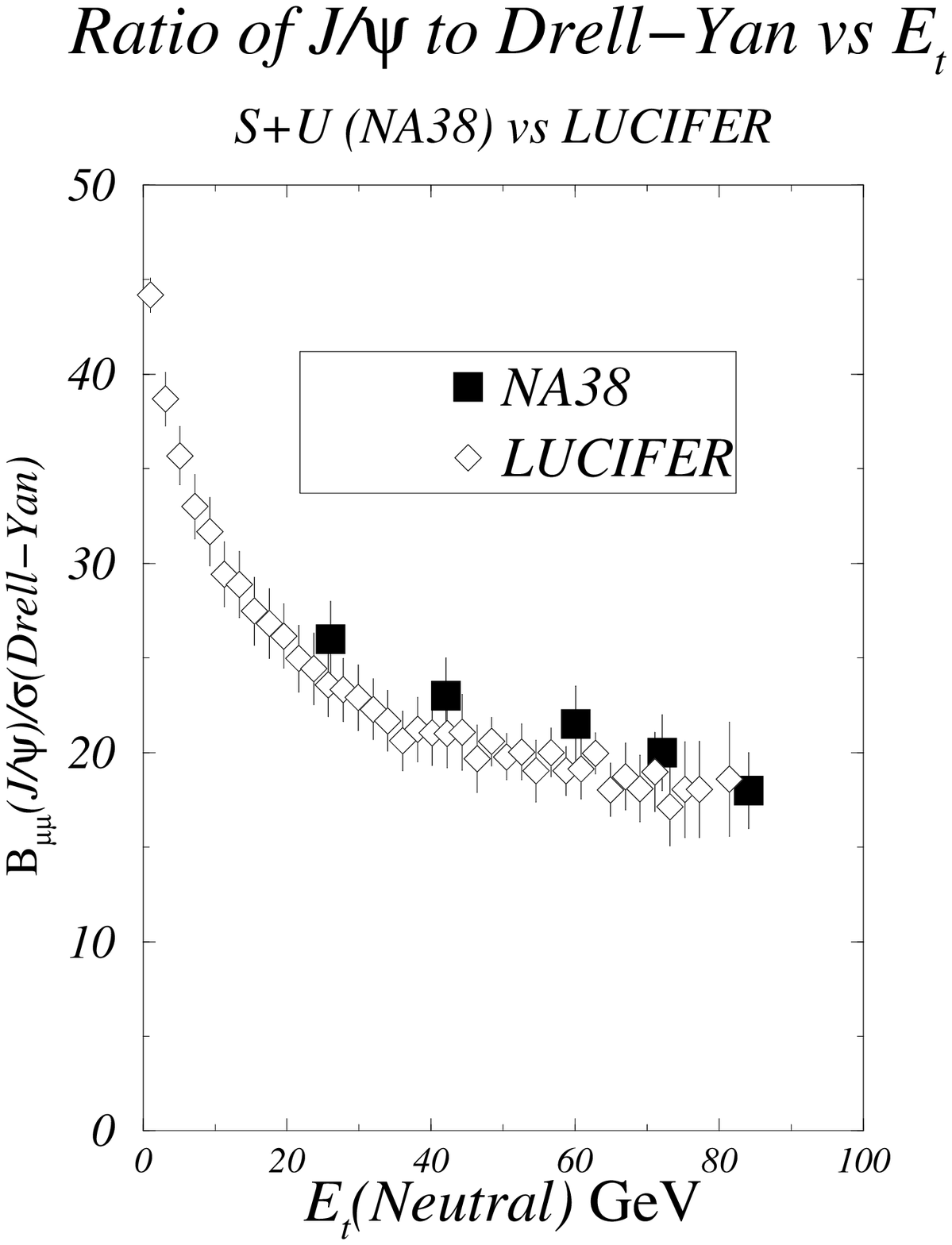} }
\caption[]{Comparison between the cascade and NA38 transverse energy 
dependence for J/$\psi$, for S+U.One notes that the pseudo rapidity
range used for the neutral $E_t$ in the case of $S+U$ is broader,
$\eta=1.7$--$4.1$, than in NA50. The calculated $E_t$ spectrum shown was
obtained with no scale factor.}
\label{fig:SUcentral}
\end{minipage}
\end{figure}

Perhaps the most striking features of the NA50 \cite{NA50b} measurements are
contained in their plot of J/$\psi$ suppression vs $E_t$. Unlike the existing
Glauber calculations of transverse energy the cascade provides a built in
$E_t$ scale, which does not necessarily agree exactly with the experimental
determination.   NA50 plots for J/$\psi$ and Drell-Yan show $E_t$ from neutral
energy  within the pseudo-rapidity range $\eta=1.1$--$2.3$. To establish a 
calibration from LUCIFER II we first refer to their earlier Pb+Pb 
results \cite{NA49} using a more central rapidity range $\eta=2.1$--$3.4$,
and including both electromagnetic (neutral) and hadronic calorimeters to
estimate $E_t$. This comparison is shown in Fig(\ref{fig:ET-NA49})
\cite{NA49}, and indicates that LUCIFER II, with standard parameters
\cite{LUCIFERII}, provides a reasonable representation of the measurements.
The small discrepancy between cascade and experimental endpoints, some
$10-15\%$, should be kept in mind when examining the NA50 charmonium data.

\begin{figure}[htb]
\begin{minipage}[t]{65mm}
	   \epsfysize = 65 mm
	   \centerline{ \epsfbox{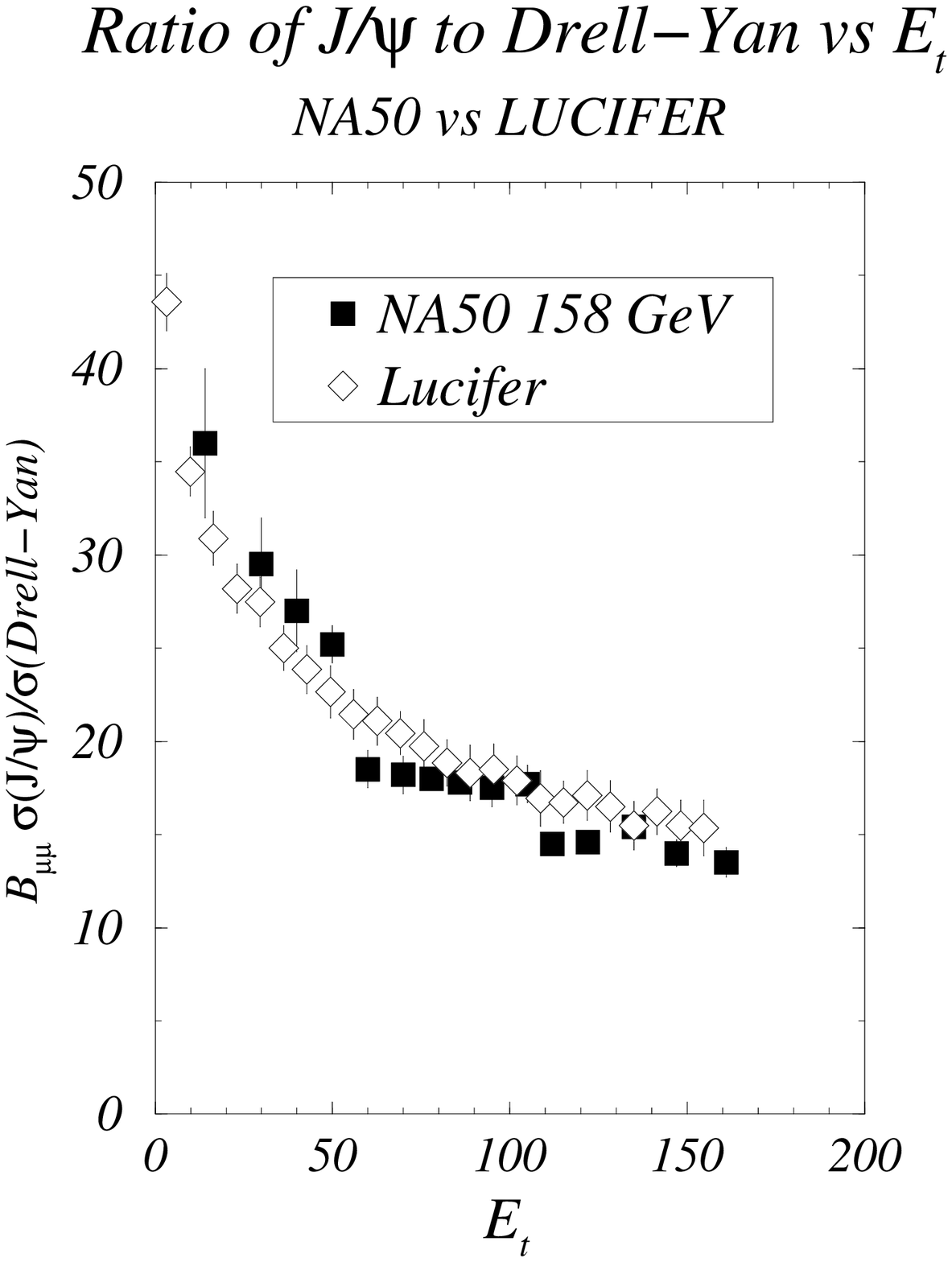} }
\caption[]{Comparison between the cascade and NA50 transverse energy 
dependence for J/$\psi$. There are no discontinuities, of course, in the
LUCIFER yields, but the general shape is reproduced. See the text for a
discussion of the $E_t$ scale.}
\label{fig:PbPbcentral}
\end{minipage}
\end{figure}

Figures (\ref{fig:SUcentral}) and (\ref{fig:PbPbcentral}) display the results
of simulations for the two massive ion-ion collisions. The magnitudes use the
calculated survival probabilities, normalised by the $pp$ or $p+D$
experiments.  The rather low $E_t$ value at which the measured J/$\psi$
suppression becomes pronounced more or less obtains in the simulation, and
the low level of J/$\psi$'s for higher $E_t$ is reproduced. The theoretical
errors shown are only indicative of the monte-carlo simulation and given the
normalisation to Drell-Yan ratios for nucleon-nucleon should include some
further systematic normalisation error. Appropriate integrals of the
$E_t$ plots  are, however, consistent with the minimum bias results in Figure
(\ref{fig:jpsi-minbias}) for both $S+U$ and $Pb+Pb$.

The results reinforce the perception already created by the comparison with
the minimum bias data. The hadronic two-step cascade is capable of describing
the charmonium yields for J/$\psi$ and $\psi'$ as well. The beginning of strong
suppression in J/$\psi$ at close to peripheral collisions, i.~e. at low
$E_t$, is a reflection of the role the heavier charmonium states play. The
scale used for central $S+U$ is just that obtained from the cascade. For
$Pb+Pb$, where NA50 used a more peripheral cut to obtain the neutral energy
the theoretical cutoff is close to $125$ GeV, somewhat lower than say that
quoted by the experiment \cite{NA50a,NA50b}, and a scaling factor of $\sim
1.25$ has been applied to the theoretical transverse energy, justified by
subsquent discussions with the experimentalists \cite{Kluberg}. The original
theoretical scale is, however, closer to that inferred from the NA49
measurements Reference \cite{NA49}. Of this factor of $1.25$, some $10-15\%$
is, as we indicated, attributable to differences in the NA49 $E_t$ calculated
and measured scales.

\section{Conclusions}

It appears that a conventional hadronic explanation of the minimum bias
and central J/$\psi$ and $\psi'$ suppression in A+B collisions is
possible. This is accomplished here with a cascade, not specially tuned for
the charmonium sector alone, but consistent with soft energy loss processes,
both meson and proton spectra, and Drell-Yan
data. There are parameters in the model, notably the meson formation time,
the generic meson decay constant, and certainly the elements of the
charmonium reaction matrix. However, we have not made use of all of this
freedom in obtaining the main results. Indeed, the only parameter not
constrained by independent information is the charmonium breakup
cross-section on mesons. The variation of suppression probability for a 
reasonable range of this variable, as seen in Figure (\ref{fig:comovers}), 
is very small. 

Some authors \cite{Kharzeev} suggest theoretically that the J/$\psi$ total
cross-section on mesons must be drastically smaller than the $\sim 4$ mb we
use for breakup, but this argument is disputed by others \cite{Huefner2}, and
it seems unlikely. In any case breakup on mesons occurs at quite high
relative energies where the very small cross-sections suggested by these
authors, even if valid, would surely not obtain.  Our direct experimental
knowledge of the total and partial cross-sections, including any energy
dependence, for J/$\psi$ or other charmonium mesons on the lower mass mesons
is of course very limited.

What then has been learned about excited, dense, nuclear matter from the
reduction in J/$\psi$'s? Our earlier calculations \cite{LUCIFERII} for a
broader range of processes, suggested that very high baryonic and mesonic
energy densities were achieved in central Pb+Pb interactions, $\rho_B\sim
4$--$5$ GeV/(fm)$^3$ and $\rho_E\sim 3$ GeV/(fm)$^3$ respectively and that
these densities persist for quite long times $\tau \sim 3$--$5$ fm/c, in the
c.m. frame. Thus appropriate conditions for possible `plasma' creation exist in
the most massive collision. This matter density has been sensed in the
theoretical comover breakup of charmonium, both J/$\psi$ and $\psi'$, more so
for Pb+Pb than for S+U. But should our model stand the test of time, and it
has explained a good portion of existing data at the SPS, then the case for a
non-conventional explanation is hard to establish as yet.  It is always of
course possible that partons actually are playing a less passive role than
portrayed by hadronic modeling and that especially high gluon densities are 
achieved in the initial phase.  Our simulations do not rule out the creation
of some form of partonic matter in ion collisions at SPS energies. They only
make the necessity thereof less compelling. A microscopic, non-hadronic
treatment of the internal strusture of the charmonium states might alter the
entire picture, and may be necessary at higher energies.

\begin{ack}
We are grateful to Louis Kluberg for several enlightening discussions. This 
manuscript has been authored under US DOE grants NO. DE-FG02-93ER407688
and DE-AC02-76CH00016.
\end{ack}

\end{document}